\documentclass[letterpaper]{article}

\usepackage[margin=1in]{geometry}
\usepackage[utf8]{inputenc}
\usepackage{mathtools}
\usepackage{amsmath}
\usepackage{amsthm}
\usepackage{amsfonts}       
\usepackage{nicefrac}       
\usepackage{xcolor}
\usepackage{soul}
\usepackage{dsfont}
\usepackage{natbib}
\usepackage{graphicx}
\usepackage{subfig}
\usepackage{multirow}
\usepackage{algorithm}
\usepackage{algorithmic}
\usepackage{comment}

\graphicspath{ {./figures/} }
\newcommand{\real}{\mathbb{R}}
\newcommand{\rE}{\mathbb{E}}
\newcommand{\rP}{\mathbb{P}}

\newcommand{\bu}{\boldsymbol{u}}

\newcommand{\bF}{\boldsymbol{F}}
\newcommand{\bL}{\boldsymbol{L}}
\newcommand{\s}{\boldsymbol{s}}
\newcommand{\m}{\boldsymbol{m}}

\newcommand{\Q}{\boldsymbol{Q}}

\newcommand{\X}{\mathcal{X}}

\newcommand{\ep}{\epsilon}
\newcommand{\dep}{\dot{\epsilon}}
\newcommand{\bep}{\boldsymbol{\epsilon}}
\newcommand{\dbep}{\boldsymbol{\dot{\epsilon}}}
\newcommand{\al}{\alpha}
\newcommand{\be}{\beta}
\newcommand{\ga}{\gamma}
\newcommand{\bga}{\boldsymbol{\gamma}}
\newcommand{\dga}{\dot{\gamma}}
\newcommand{\blm}{\boldsymbol{\lambda}}
\newcommand{\dblm}{\boldsymbol{\dot{\lambda}}}
\newcommand{\lm}{\lambda}
\newcommand{\dlm}{\dot{\lambda}}
\newcommand{\md}{\boldsymbol{\sigma}_{M}}

\newcommand{\bP}{\boldsymbol{P}}
\newcommand{\siginf}{{\sigma}_{\alpha}^{\infty}}

\newcommand{\strain}{\varepsilon}

\makeatletter
\newcommand{\printfnsymbol}[1]{
  \textsuperscript{\@fnsymbol{#1}}}

\makeatother

\title{Hierarchical multiscale quantification of material uncertainty}
\author{Burigede Liu~\footnote{Division of Engineering and Applied Science, California Institute of Technology, Pasadena, CA 91125, United States}~~\footnote{These authors contribute equally.}~, Xingsheng Sun\printfnsymbol{1}\printfnsymbol{2}, Kaushik Bhattacharya\printfnsymbol{1}, Michael Ortiz\printfnsymbol{1}}
\date{February 2021}

\begin{document}

\maketitle

\begin{abstract}

The macroscopic behavior of many materials is complex and the end result of mechanisms that operate across a broad range of disparate scales. An imperfect knowledge of material behavior across scales is a source of epistemic uncertainty of the overall material behavior. However, assessing this uncertainty is difficult due to the complex nature of material response and the prohibitive computational cost of integral calculations. In this paper, we exploit the multiscale and hierarchical nature of material response to develop an approach to
quantify the overall uncertainty of material response without the need for integral calculations. Specifically, we bound the uncertainty at each scale and then combine the partial uncertainties in a way that provides a bound on the overall or integral uncertainty. The bound provides a conservative estimate on the uncertainty. Importantly, this approach does not require integral calculations that are prohibitively expensive. We demonstrate the framework on the problem of ballistic impact of a polycrystalline magnesium plate. Magnesium and its alloys are of current interest as promising light-weight structural and protective materials. Finally, we remark that the approach can also be used to study the sensitivity of the overall response to particular mechanisms at lower scales in a materials-by-design approach.
\end{abstract}{}


\section{Introduction}

The macroscopic behavior of many materials is complex and the end result of mechanisms that operate across a broad range of disparate scales \citep{p_book_01}. The mesoscopic scales both filter (average) and modulate (set the boundary conditions or driving forces for) the mechanisms operating at lower scales, which establishes a functional hierarchy among mechanisms. The complexity of the material response is often a main source of uncertainty in engineering applications, a source that is epistemic in nature and traceable to imperfect knowledge of material behavior across scales. This epistemic uncertainty often renders deterministic analysis of limited value. Instead, integral uncertainties must be carefully quantified in order to identify adequate design margins and meet design specifications with sufficient confidence. However, the direct estimation of integral material uncertainties entails repeated calculations of integral material response aimed at determining worst-case scenarios at all scales resulting in the largest deviations in microscopic behavior. Such integral calculations are almost always prohibitive and well beyond the scope of present-day computers.

The complexity of materials response also poses a challenge to the development of new material systems and the optimization of properties through a `materials-by-design' approach \citep{olson2000designing}. In this approach, one seeks to `design' materials with desired properties by targeting individual mechanisms at lower scales.  However, while one can affect individual mechanisms and assess the response at a particular scale, for example by adding solutes to affect dislocation kinetics and assessing it using molecular dynamics \citep{kohler2004atomistic}, it is extremely challenging to understand how these changes percolate through the hierarchy.  In other words, the complexity makes it extremely difficult to understand the sensitivity of the overall material response to individual mechanisms.  Once again, direct estimation leads to computational problems that are well beyond the scope of present-day computers.  While our current work concerns uncertainty, there is a close connection between these issues.

Conveniently, the very multiscale and hierarchical nature of material response itself can be exploited for purposes of uncertainty quantification. We recall that multiscale modeling is fundamentally a 'divide-and-conquer' paradigm whereby the entire range of material behaviors is divided into a hierarchy of length scales \citep{Ortiz:2001}. The relevant unit mechanisms are then identified, namely, physical mechanisms that are irreducible and operate roughly independently: two mechanisms that are tightly coupled should be considered as a single unit mechanism. In this hierarchy, the unit mechanisms at one scale represent averages of unit mechanisms operating at the immediately lower length scale. This functional relation introduces a partial ordering of mechanisms that defines a directed graph. The nodes of the graph are the unit mechanisms, the root represents the integral macroscopic behavior of the solid and the edges define the upward flow of information from the leaves to the root of the graph.

\begin{figure}[!ht]
\centering
\includegraphics[width = 5in]{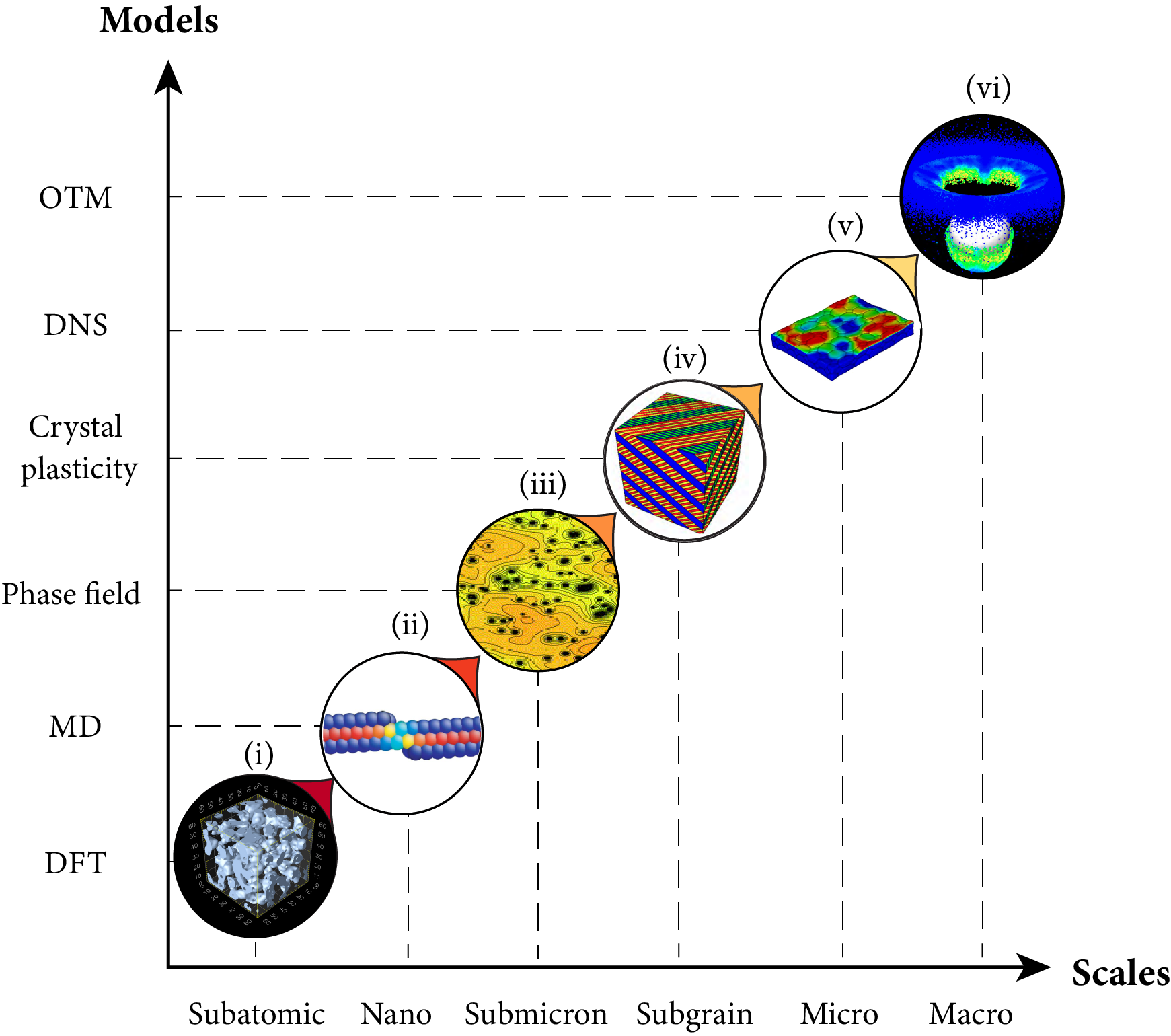}
\caption{Example of multiscale hierarchy for modeling strength in metals. Overlaid are examples of models or calculations of several of the unit mechanisms, from bottom to top: i) density functional theory (DFT) calculations of the equation of state (EoS) in Ta \citep{Miljacic:2011}; ii) molecular dynamics (MD) calculations of kink mobility in Ta \citep{Wang:2001, Segall:2001, kang2012singular}; iii) phase-field simulations of dislocation dynamics and forest hardening \citep{koslowski2002a}; iv) lamination construction for sub-grain dislocation structures \citep{ortiz1999a}; v) direct numerical simulation (DNS) of polycrystals \citep{zhao2004a}; vi) Optimal Transportation Meshfree (OTM) simulations of ballistic perforation \citep{li2010a}. }
\label{URNvuL}
\end{figure}

A representative hierarchy for modeling strength in metals is shown in Fig.~\ref{URNvuL} by way of example. A number of fundamental properties, such as the equations of state describing the compressibility of the material, the elastic moduli and heat capacity, can be calculated from first principles at the quantum-mechanical level up to high pressures and temperatures. Some transport properties, such as viscosity and thermal conductivity, can also be characterized from first principles, e.~g., within the Green-Kubo formalism~\citep{green1954markoff, kubo1957statistical}. At the nanoscale, the properties of individual lattice defects, such as vacancies and dislocations, come into focus \citep{leibfried2006point}. Such properties include dislocation core energies, kink mobilities, lattice friction, short-range dislocation-dislocation interactions, vacancy formation and migration energies, grain-boundary energies, and others. The sub-micron scale is dominated by dislocation dynamics \citep{messerschmidt2010dislocation}, i.~e., the cooperative behavior of large dislocation ensembles, and, in the case of ductile fracture, void nucleation and cavitation. The sub-grain scale is characterized by the formation of highly-structured dislocation patterns and, in the case of metals undergoing solid-solid phase transitions, martensitic structures \citep{bhattacharya2003microstructure}. This intermediate scale is important, as it underlies scaling properties such as Hall-Petch scaling, or the inverse relation between strength and the square-root of the grain size \citep{hall1951deformation,petch1953cleavage}. At the microscale, ductile fracture is characterized by plastic void growth and coalescence. Finally, the macroscopic response of polycrystalline metals represents the effective behavior of large ensembles of grains (cf., e.~g., \citep{hutchinson1970elastic, wei2004grain} for reviews).

In this work we show how, in materials for which a multiscale hierarchy is well defined, the quantification of integral uncertainties can be reduced to the analysis of each unit mechanism in turn and the propagation of unit uncertainties across the multiscale hierarchy according to an appropriate measure of interaction between the unit mechanisms. In particular, \emph{no integral calculation is required at any stage of the analysis}. We specifically follow the approach of \citet{topcu2011rigorous}, which supplies rigorous upper bounds of integral uncertainties for hierarchies of interconnected subsystems through a systematic computation of moduli of continuity for each individual subsystem. The moduli of continuity supply just the right measure of interaction between the subsystems enabling the propagation of uncertainties across the hierarchy. The resulting uncertainty bounds are rigorous, i.~e., they are sure to be conservative and result in safe designs; they become sharper with an increasing number of input variables (concentration-of-measure phenomenon~\citep{talagrand1996new}); they do not require differentiability of the subsystem response functions and account for large deviations in the response; and only require knowledge of ranges of the input parameters and not their full probability distribution, as is the case for Bayesian methods \citep{dashti2011uncertainty}. In addition, the computation of the bounds is non-intrusive and can be carried out using existing deterministic models of the subsystems and external scripts.

We specifically aim to assess this hierarchical multiscale approach to uncertainty quantification (UQ) by means of an example concerned with the ballistic impact of an elastic-plastic magnesium plate struck by a heavy rigid ball. We adopt as quantity of interest the maximum backface deflection of the plate. We consider two scales of material response: a macromechanical scale characterized by the Johnson-Cook constitutive model \citep{johnson1983constitutive}; and a micromechanical scale in which the polycrystalline structure of the material and its behavior at the single-crystal level are taken into account. For simplicity, we compute polycrystalline averages by means of Taylor averaging~\citep{t_jim_38} assuming an isotropic initial texture. The single-crystal plasticity extended to include twinning follows \citet{chang2015variational}.  We estimate ranges of parameters for the single-crystal plasticity model based on experimental data compiled from a number of sources. Calculations are carried out using the commercial finite-element package LS-DYNA~\citep{hallquist2007ls} on a single converged mesh. The analysis mirrors conventional design testing for impact resistance~\citep{standard20080101}, wherein performance is evaluated relative to a targeted set of characterized impact conditions. The hierarchical multiscale UQ protocol is implemented using the DAKOTA Version $6.12$ software package~\citep{adams2020dakota} of the Sandia National Laboratories.

It bears emphasis that the material model used in the calculations is intended for purposes of demonstration of the methodology and not as an accurate model of material behavior. This proviso notwithstanding, the calculations show that the integral uncertainties determined by the hierarchical multiscale UQ approach are sufficiently tight for use in engineering applications. The analysis also sheds light on the relative contributions of the different unit mechanisms to the integral uncertainty and the dominant propagation paths for uncertainty across the model hierarchy.

\section{Methodology}
\label{sec:method}

We start with a brief review of the uncertainty quantification theory for hierarchical systems \citep{lucas2008rigorous, topcu2011rigorous, sun2020rigorous} that provides the basis for the application to multiscale material modelling pursued in this work.

\subsection{McDiarmid's inequality}
We begin by considering a single-scale, or monolithic, system with uncertain inputs $x = (x_1, ..., x_N)$ and a performance measure $y \in \real$. We assume that the inputs are independent, though not necessarily identically distributed, random variables defined on a bounded set $\X \subset \real^N$. Let ${F}: \X \mapsto \real$ be the response function that maps $x$ to $y$. Define the {\sl diameter} of ${F}$ with respect to input $x_i$, $1\leq i \leq N$, as
\begin{equation}
\label{eq:bounded_difference}
    d_{F,i}
    =
    \sup_{(\hat{x}_i, x_i), (\hat{x}_i, x_i')\in \X}
    \big|{F}(\hat{x}_i,x_i)-{F}(\hat{x}_i, x_i')\big| ,
\end{equation}
where we write
\begin{equation}
    \hat{x}_i
    =
    (x_1,\ldots,x_{i-1},x_{i+1},\ldots,x_N) .
\end{equation}
Given a performance threshold $y_c \in \real$, with failure occurring when $y \geq y_c$, the McDiarmid's inequality \citep{doob1940regularity} gives the following upper bound on the probability of failure
\begin{equation}
\label{eq:MicDiarmid}
    \rP[y \geq y_c] \leq \text{exp}\bigg(-2\frac{M^2}{U^2}\bigg),
\end{equation}
where $M = \text{max}(0,y_c-\rE [y])$ is the design margin and
\begin{equation}\label{SKnLZH}
    U = d_F = \bigg(\sum_{i=1}^N d_{F,i}^2 \bigg)^{1/2}
\end{equation}
is the system uncertainty. Thus, the system is certified if, for a given probability-of-failure tolerance $\epsilon > 0$,
\begin{equation}\label{eq:PoF}
    \text{exp}\bigg(-2\frac{M^2}{U^2}\bigg) \leq \epsilon ,
\end{equation}
or equivalently,
\begin{equation}
\label{eq:Confident factor}
\text{CF} = \frac{M}{U} \geq \sqrt{\text{log}\sqrt{\frac{1}{\epsilon}}},
\end{equation}
where $\text{CF}$ is a confidence factor in the design \citep{sharp2003qmu}.

We note that Eq.~(\ref{eq:MicDiarmid}) provides a rigorous probability-of-failure upper bound for the material and, therefore, sets forth a conservative certification criterion. Importantly, no a-priori assumption on the probability distribution of the input variables, or {\sl prior}, is required. The bound is uniquely determined by the design margin $M$ and the system uncertainty $U$, as unambiguously quantified by the system diameter $d_F$, which are both prior-free.

\subsection {Hierarchical Uncertainty Quantification}

As noted in the introduction, the macroscopic behavior of many material systems is the result of mechanisms over multiple scales whose functional dependencies define a graph. Conveniently, this graph structure can be exploited to divide the material response into interconnected unit mechanisms, or subsystems, and estimate integral uncertainties of the entire system from a quantification of uncertainties for each subsystem and an appropriate measure of interaction between the subsystems. We specifically follow the approach of \citet{topcu2011rigorous}, which we briefly summarize next.

\begin{figure}
\begin{center}
\includegraphics[width=4.50in]{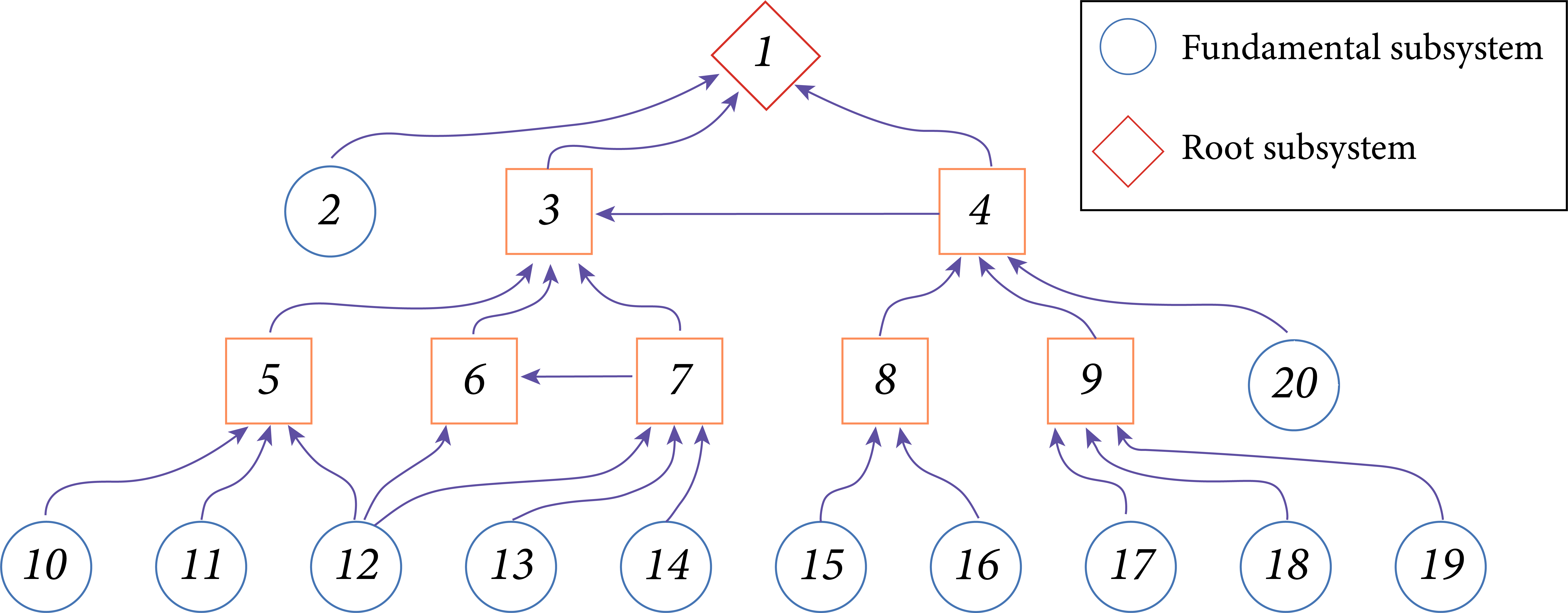}
\end{center}
\caption{Graph representation of the input-output relations between the subsystems of a modular system. The nodes of the graph represent the subsystems of the system. An arrow indicates that the outputs of the subsystem at the beginning of the arrow are among the inputs of the subsystem at the end of the arrow. The subsystems represented by circular boxes do not take input from other systems and are referred to as {\sl fundamental}. The system has one single {\sl root} subsystem, i.~e., a subsystem that does not feed into any other subsystem and through which the system output takes place.} \label{fig:houmanGraph}
\end{figure}

We consider hierarchical or modular systems representable by an oriented graph $G(V,E)$ whose nodes $V$ are the subsystems and whose edges $E$ are the interfaces between the subsystems, Fig.~\ref{fig:houmanGraph}. Specifically, the graph contains an oriented edge from $b$ to $a$ if the state of the subsystem $a$ depends on the state of the subsystem $b$, i.~e., if the outputs of subsystem $b$ are contained among the inputs of system $a$. We then say that $a$ is an {\sl ancestor} of a node $b$, denoted $a \prec b$, and that $b$ is a {\sl descendant} of $a$, denoted $b \succ a$. In order to avoid circular dependencies, we assume that $G(V,E)$ is {\sl acyclic}, i.~e., it contains no closed-loop paths. The {\sl fundamental} subsystems are those that take input from no other subsystems. We assume that the system contains a single {\sl integral} subsystem that does not feed into any other subsystem. Thus, the fundamental subsystems and the integral subsystem are the leaves $V_L$ and the root $R$ of the graph $G(V,E)$, respectively.

Suppose that the response of subsystem $a \in V$ is characterized by a function $F_a : \mathcal{X}_a \to \mathcal{Y}_a$ that maps input parameters $X_a \in \mathcal{X}_a$ to outputs $Y_a \in \mathcal{Y}_a$. If $a$ is an ancestor of $b$ in the graph $G(V,E)$ then $\mathcal{Y}_b$ is a subspace of $\mathcal{X}_a$. The space of inputs of the system $\mathcal{X}$ is the Cartesian product of the input spaces of the fundamental subsystems, i.~e., $\mathcal{X} = \prod_{a\in V_L} \mathcal{X}_a$. Likewise, the space of outputs of the system is $\mathcal{Y} = \mathcal{Y}_R$. For all subsystems other than the fundamental ones, $a \not\in V_L$, we have the relation $\mathcal{X}_a = \prod_{b \succ a} \mathcal{Y}_b$, i.~e., the input space of subsystems $i$ is the Cartesian product of the output spaces of all its descendants. We note that non-fundamental subsystems could in principle have inputs of their own, not provided by any descendant subsystem. We accommodate such cases within the present framework simply by adding a fundamental subsystem whose response function is the identity mapping and which supplies the requisite additional inputs.

\begin{algorithm}[H]
\caption{Hierarchical System Evaluation}
\label{alg:Evaluation}
\begin{algorithmic}
    \REQUIRE Graph $G(V,E)$; leaves $V_L$; root $R$; response functions $F_a : \mathcal{X}_a \to \mathcal{Y}_a$, $a \in V$; input $X\in \mathcal{X}$.
    \STATE i) Initialize: $V_0 = V_L$, $k=0$.
    \STATE ii) Reset: $V_{k+1} = \{a\in V :\ b \in \cup_{l=0}^k V_l,\ \forall b\succ a\}$.
    \FORALL {$a \in V_{k+1}$}
        \STATE Compute $X_a = \big( F_b(X_b),\ b\succ a \big)$.
    \ENDFOR
    \IF {$V_{k+1} = \{R\}$}
        \STATE return $Y=F_R(X_R)$, {\bf exit}.
    \ELSE
        \STATE $k \leftarrow k+1$, {\bf goto} ii).
    \ENDIF
\end{algorithmic}
\end{algorithm}

The function $F : \mathcal{X} \to \mathcal{Y}$ that describes the response of the integrated system can be evaluated recursively by means of Algorithm~\ref{alg:Evaluation}. The algorithm sets forth an ``information wave" through the graph that propagates information from the leaves to the root. Thus, in the example of Fig.~\ref{fig:houmanGraph}, the sequence of active subsystems is $V_0=\{2,10,11,12,13,14,15,16,17,18,19,20\}$, $V_1=\{5,7,8,9\}$, $V_2=\{6,4\}$, $V_3=\{3\}$ and $V_5=\{1\}$. Correspondingly, the sequence of subsystem outputs is $(Y_5, Y_7, Y_8, Y_9)$, $(Y_6, Y_4)$, $Y_3$ and $Y=Y_1$.

Let $\{V_k, k=0,\dots,N\}$ be the sequence of nodal sets generated during the evaluation of the integrated response function $F(X)$, with $V_0 = V_L$ and $V_N = \{R\}$. Define $\mathcal{X}_k = \prod_{a\in V_k} \mathcal{X}_a$ and $\mathcal{Y}_k = \prod_{a\in V_{k+1}} \mathcal{X}_a$, i.~e., the combined sets of inputs and outputs for each iteration. We note that $\mathcal{X} = \mathcal{X}_0$, $\mathcal{Y}_k = \mathcal{X}_{k+1}$. Without loss of generality we may assume that $\mathop{\rm dim}\mathcal{Y}_N = \mathop{\rm dim}\mathcal{Y} = 1$. Define further $F_k : \mathcal{X}_k \to \mathcal{Y}_k$ as $F_k(X_k) = (F_a(X_a), a\in V_k)$, $X_k\in \mathcal{X}_k$, i.~e., as the forward map for iteration $k$. By these definitions and reorganization of the data, we have the composition rule
\begin{equation}
    F = F_N \circ\cdots \circ F_0.
\end{equation}

Conveniently, the propagation of uncertainty under composition is controlled by the moduli of continuity of the response functions. Recall that, given a function $f: \mathbb{R}^n \to \mathbb{R}^m$, a real number $\delta > 0$ and a subset $A \subset \mathbb{R}^n$, the modulus of continuity $\omega_{ij} (f, \delta, A)$ of $f_i(x)$ with respect to $x_j$ over $A$ is defined as \citep{efimov2001modulus, steffens2006constructive}
\begin{equation} \label{eq:MoC}
\begin{split}
    &
    \omega_{ij}(f,\delta,A)
    = 
    \sup
    \{|f_i(x) - f_i(x')| \ : \
    x, x' \in A,\
    x_k=x'_k\;\text{for}\; k\not=j,\
    |x_j - x'_j| \leq \delta\}.
\end{split}
\end{equation}
Thus, $\omega_{ij} (f, \delta, A)$ measures the variation of the function $f_i(x)$ over $A$ when the variable $x_j$ is allowed to deviate by less than $\delta$. We note that this component-wise definition of the modulus of continuity does not require the range or image of the function $f$ to be a normed space. This is important in practice, since the inputs and outputs of subsystems often comprise variables measured in different units which belong to vector spaces with no natural norm.
Consider now two functions $f: A \subset \mathbb{R}^n \to \mathbb{R}^m$ and $g: B \subset \mathbb{R}^m \to \mathbb{R}^p$, with $B$ a hyper-rectangle such that $f(A) \subset B$, and let $\delta > 0$. Let $g\circ f: A \subset \mathbb{R}^n \to \mathbb{R}^p$ be the composition of the $f$ and $g$, i.~e., $(g\circ f)(x) = g(f(x))$. Then, we have \citep{topcu2011rigorous}
\begin{equation}\label{eq:MoC5}
    \omega_{ij}(g\circ f,\delta, A)
    \leq
    \sum_{k=1}^m \omega_{ik}(g,\omega_{kj}(f,\delta,A),B).
\end{equation}
This inequality shows that the moduli of continuity of a composite function $g\circ f$ can be estimated conservatively from the moduli of continuity of the individual functions.

\begin{algorithm}[H]
\caption{Hierarchical Uncertainty Quantification}
\label{alg:Solver}
\begin{algorithmic}
    \REQUIRE Graph $G(V,E)$; leaves $V_L$; root $R$; response functions $F_a : \mathcal{X}_a \to \mathcal{Y}_a$, $a \in V$; set $A\subset \mathcal{X}_0 = \mathcal{X}$.
    \FORALL {$j=1,\dots,{\rm dim}\mathcal{X}_0$}
        \STATE Compute: $D_j = \sup\{|x_j-x'_j| :\ x, x'\in A,\ x_k=x_k',\text{ for } k\neq j\}$.
        \STATE Compute: $D_{ij}^{(0)} = \omega_{ij}(F_0,D_j,A), \ i = 1,\dots,\mathop{\rm dim}\mathcal{Y}_0$.
    \ENDFOR
    \FORALL {$k=1,\dots,N$}
        \STATE Find hyper-rectangles $B_k \subset \mathcal{X}_k$ containing $(F_{k-1} \circ\cdots\circ F_0)(A)$.
        \STATE Compute: $D_{ij}^{(k)} = \sum_{l=1}^{\mathop{\rm dim}\mathcal{X}_k} \omega_{il}(F_k,D_{lj}^{(k-1)},B_k), \ i = 1,\dots,\mathop{\rm dim}\mathcal{Y}_k$.
    \ENDFOR
    \STATE {\bf return} $\{D_{F,i} = D_{1i}^{(N)},\ i=1,\dots,{\rm dim}\mathcal{X}\}$.
\end{algorithmic}
\end{algorithm}

This property of the moduli of continuity in turn enables uncertainties of the integral system to be bounded once the uncertainties of the subsystems and their interfaces are known. The systematic application of estimate (\ref{eq:MoC5}) to the graph $G(V,E)$ is described in Algorithm \ref{alg:Solver} and results in a set of approximate diameters $\{D_{F,i},\ i=1,\dots,{\rm dim}\mathcal{X}\}$.  The fundamental theorem proven by \citet{topcu2011rigorous} is that the approximate diameters $\{D_{F,i}\}$ bound above the exact integral diameters $\{d_{F,i}\}$, Eq.~(\ref{eq:bounded_difference}), i.~e.,
\begin{equation}\label{eq:DFBound}
    d_{F,i} \leq D_{F,i} ,
    \ i=1,\dots,{\rm dim}\mathcal{X}.
\end{equation}
By the monotonicity of McDiarmid's inequality, with respect to the diameters, it follows that replacing $\{d_{F,i}\}$ by $\{D_{F,i}\}$ in (\ref{SKnLZH}) and (\ref{eq:PoF}) results in probabilities of integral outcomes and, by extension, in conservative certification criteria.

It bears emphasis that every step of Algorithm \ref{alg:Solver} requires the execution of subsystem tests only, and that at no time during the analysis an integral test is required. The sequence $D_{ij}^{(k)}$ generated by the algorithm may be regarded as a measure of uncertainty in the $i$th output variable due to the variability of the $j$th input variable after $k$ levels of operation of the system. The algorithm propagates uncertainties in the input variables associated with a leaf $i$ through possibly multiple {\it paths} connecting the node $i$ with the root $R$. The algorithm additionally identifies the path responsible for most of the uncertainty of the integral outcome with respect to the variable $i$, namely, the path with the highest flow of uncertainty.

\section{Test case: Sub-ballistic impact of magnesium plate}
\label{sec:sim}

\begin{figure}[!ht]
    \centering
    \includegraphics[width=\textwidth]{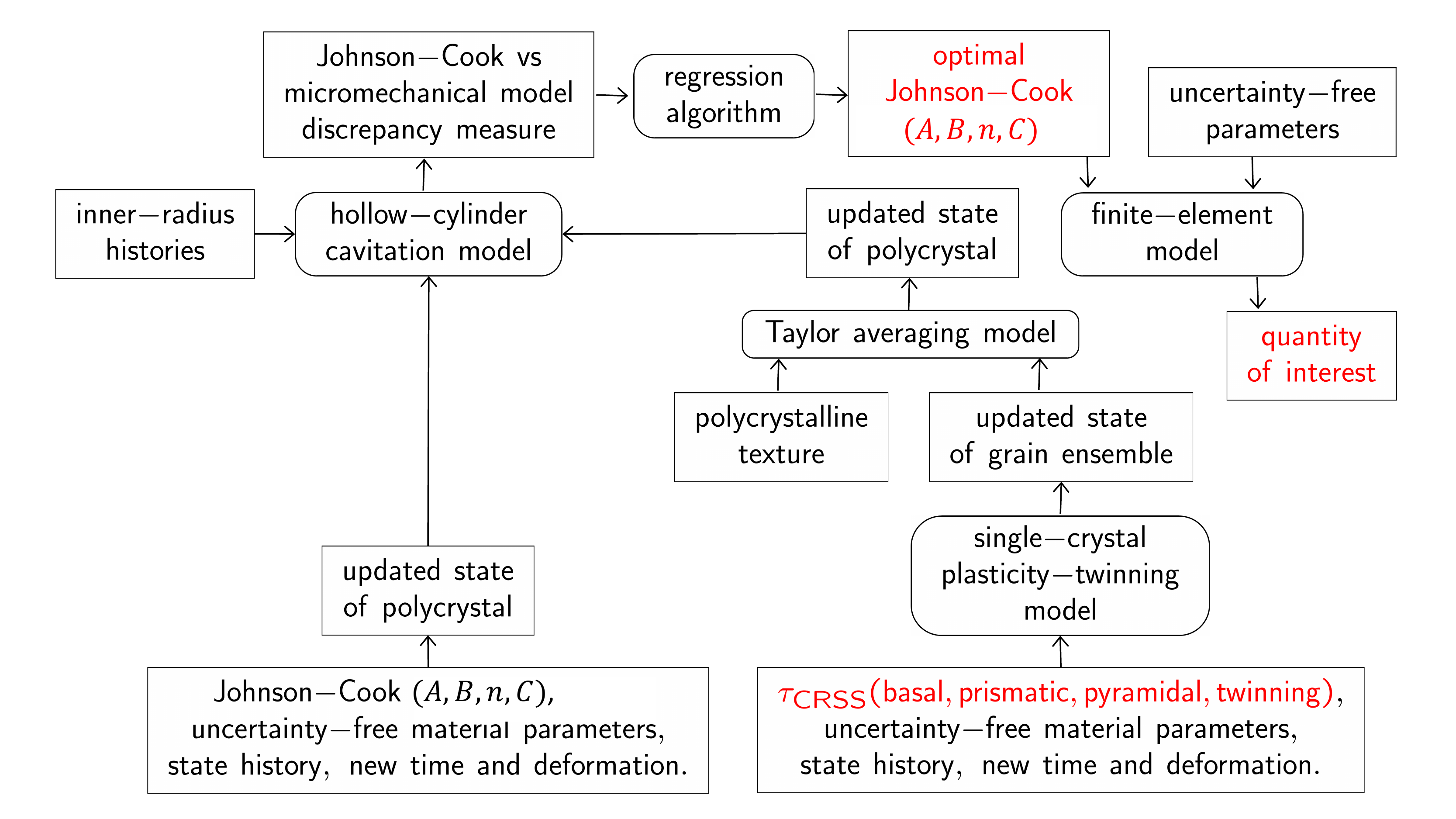}
    \caption{Graph representation of the hierarchical multiscale model of ballistic impact. Highlighted in red are the micro and mesomechanical parameters assumed to be uncertain in the UQ analysis. }
    \label{fbEJGz}
\end{figure}

As just shown in the foregoing, the integral uncertainty for a hierarchical multi-scale system can be evaluated from the moduli of continuity of the sub-system maps without integral testing. We proceed to demonstrate the feasibility of the approach by means of an example of application concerning the ballistic impact of a magnesium plate. For purposes of this demonstration, we restrict attention to three length scales as depicted in Fig.~\ref{fbEJGz}: i) the microscale, where the behavior of the magnesium, including slip and twinning, is modelled at the single crystal level; ii) the mesoscale, where the polycrystalline response is computed using Taylor averaging and the single-crystal model; and iii) the macroscopic impact problem, where the material behavior is approximated by the Johnson-Cook constitutive model and the ballistic performance of the magnesium plate is simulated using finite elements.

For purposes of illustration of the UQ methodology, we assume that all parameters are uncertainty-free save those shown in red in Fig.~\ref{fbEJGz}, namely the micromechanical critical resolved shear stresses (CRSS) for the basal slip, prismatic slip, pyramidal slip and twinning mechanisms, the mesomechanical Johnson-Cook parameters and the integral quantity of interest. These assumptions lead to the following simplified two-level hierarchical system:
\begin{equation}\label{RF3BSm}
    \mathcal{X}_0
    \stackrel{F_0}{\longrightarrow}
    \mathcal{Y}_0 = \mathcal{X}_1
    \stackrel{F_1}{\longrightarrow}
    \mathcal{Y}_R ,
\end{equation}
cf.~Fig.~\ref{fbEJGz}. Thus, the graph $V$ of the system contains three nodes, denoted $\{{\rm micro}\equiv 0,\ {\rm meso}\equiv 1,\ {\rm macro}\equiv R\}$, and
\begin{itemize}
    \item[$\mathcal{X}_0 =$] Uncertain single-crystal model parameters: (CRSS) for the basal slip, prismatic slip, pyramidal slip and twinning mechanisms.
    \item[$F_0 =$] Mapping that returns an optimal set $\mathcal{Y}_0$ of Johnson-Cook parameters $(A, B, n, C)$ for a given realization of $\mathcal{X}_0$, obtained by performing regression on stress paths computed using both the Johnson-Cook model and the single crystal model combined with Taylor averaging along selected strain paths.
    \item[$\mathcal{Y}_0 =$] $\mathcal{X}_1 =$ Optimal Johnson-Cook parameters.
    \item[$F_1 =$] Finite element model of ballistic impact of magnesium plate using the Johnson-Cook model with optimal parameters $\mathcal{X}_1$.
    \item[$\mathcal{Y}_R =$] Quantity of interest extracted from the results of the finite element calculations.
\end{itemize}

A brief description of the different subsystems is given next, followed by a specification of the maps $F_0$ and $F_1$.

%


\subsection{Micromechanical model}

\begin{figure}[!ht]
\centering
\includegraphics[width=5.0in]{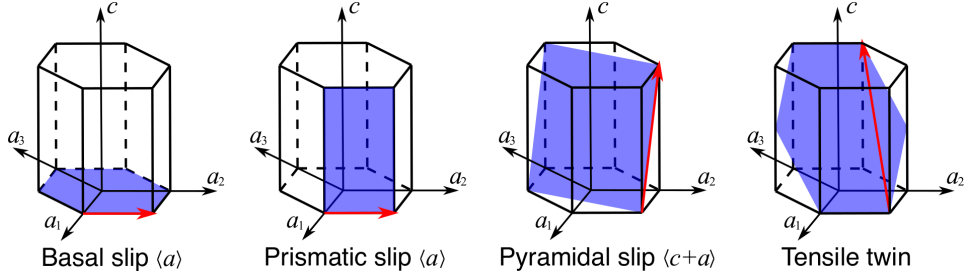}
\caption{Schematic view of the slip and twin systems of magnesium considered in the present work. Blue planes represent the slip/twin planes, while red arrows represent the corresponding Burgers vector/twinning shear direction.}
\label{fig:sliptwin}
\end{figure}

We specifically model single-crystal magnesium within the framework of finite-deformation single crystal plasticity extended to include twinning following \citet{chang2015variational}. The reader is referred to the original publication for details of the model. The model accounts for basal slip, prismatic slip, pyramidal slip and tensile twinning, cf.~Fig.~\ref{fig:sliptwin}. The entire collection of parameters of the model is listed in Table~\ref{tab:param_crystal}, including the values of the parameters determined by \citet{chang2015variational}. For each slip system $\alpha$, $\siginf$ is the ultimate strength, $h_\al$ is the self-hardening modulus, $m_\al$ is the slip rate sensitivity exponent, $\dga_{0,\al}$ is a reference slip rate and $h_{ij}$ are off-diagonals latent hardening moduli. For twinning, $h_\be$ is the self-hardening modulus, $m_\be$ is the rate-sensitivity exponent, $\dlm_{0,\be}$ is a reference twin volume-fraction rate, $\ga_t$ denotes the twin strain and $k_{ij}$ are interaction moduli. In addition, the elasticity is assumed to be isotropic with Lam\'e constants $\lambda_e$ and $G$.

\begin{table}[!ht]
\centering
\caption{Parameters of the single-crystal magnesium model.}
\begin{tabular}{l l l l}
\hline
\hline
& Parameter & Value & Unit\\
\hline
\multirow{5}{*}{Basal $\langle a \rangle$}
& $h_\al$ & 7.1 &GPa\\
& $\siginf$ & 0.7 &MPa\\
& $h_{ij}$ & 0 &MPa \\
& $m_\al$ & 0.05 & -\\
& $\dga_{0,\al}$ & 1.0 &$\text{s}^{-1}$\\
\hline
\multirow{5}{*}{Prismatic $\langle a \rangle$}
& $h_\al$ & 40 &GPa\\
& $\siginf$ & 170 &MPa\\
& $h_{ij}$ & 20 &MPa \\
& $m_\al$ & 0.05 & -\\
& $\dga_{0,\al}$ & 1.0 &$\text{s}^{-1}$\\
\hline
\multirow{6}{*}{Pyramidal $\langle c+a \rangle$}
& $h_\al$ & 30 &GPa\\
& $\siginf$ & 200 &MPa\\
& $h_{ij}$ & 25 &MPa \\
& $m_\al$ & 0.05 & -\\
& $\dga_{0,\al}$ & 1.0 &$\text{s}^{-1}$\\
\hline
\multirow{5}{*}{Tensile twin}
& $h_\be$ & 7 &MPa \\
& $k_{ij}$ & 40 &GPa \\
& $m_\be$ & 0.05 & -\\
& $\dlm_{0,\be}$ & 1.0 &$\text{s}^{-1}$ \\
& $\ga_t$ &  0.129 & -\\
\hline
\multirow{2}{*}{Elastic Lame Moduli}
& $\lambda_{e}$ & 24 &GPa \\
& $G$ & 25 &GPa \\
\hline
\end{tabular}
\label{tab:param_crystal}
\end{table}

We assume that the target plate is made of polycrystalline magnesium and compute the polycrystalline response of the polycrystal from the single crystal model just outline by means of Taylor averaging \citep{t_jim_38}. The computational cost of Taylor averaging is relatively small compared to other averaging schemes such as periodic boundary conditions \citep{geers2010multi}. In addition, \citet{chang2015variational} have demonstrated good agreement between Taylor averaging and experimental measurements when the number of the grains is sufficiently large ($>100$). In calculations we assume initial isotropic texture with grain orientation matrix drawn uniformly from the Haar measure on $SO(3)$ \citep{haar1933massbegriff}. A total of 128 grains are used in the current study for the mesoscale model.

\subsection{Johnson-Cook model}

The combination of a single-crystal model and Taylor averaging just described, together with suitable initial conditions, enables the calculation of Cauchy stress histories $\sigma(t)$ at a material point from known deformation gradient histories $F(t)$. However, such calculations are computationally intensive and cannot be carried out on-the-fly as part of large-scale finite-element calculations. Instead, in said calculations we choose to use a simpler and faster surrogate or mesomechanical model.

For definiteness, we specifically choose the Johnson-Cook model~\citep{johnson1983constitutive}
\begin{equation}
    \sigma_M \big( \epsilon_p, \dot{\epsilon}_p)
    =
    \big[A + B \epsilon_p^n \big]
    \big[1 + C \ln \dot{\epsilon}_p^* \big],
\label{eq:johnsoncook}
\end{equation}
as mesomechanical model, where $\sigma_M = \sqrt{\nicefrac{3}{2}s \cdot s}$ is the Mises stress and $s = \sigma - \nicefrac{1}{3}\text{tr}(\sigma) I$ denotes the deviatoric part of the Cauchy stress $\sigma$, $\epsilon_p$ denotes the equivalent plastic strain, $\dot{\epsilon}_p$ is the plastic strain rate and $\dot{\epsilon}_p^* = \dot{\epsilon}_p/\dot{\epsilon}_{p0}$ is a normalized plastic strain rate using reference $\dot{\epsilon}_{p0}$. The model parameters are the strength $A$, the hardening modulus $B$, the strain-hardening exponent $n$, and the rate-sensitivity modulus $C$, or $\mathcal{X}_1 = (A, B, n, C)$.

\subsection{Micro-mesomechanical parameter map}

\begin{figure}[!ht]
    \centering
    \includegraphics[height=5cm]{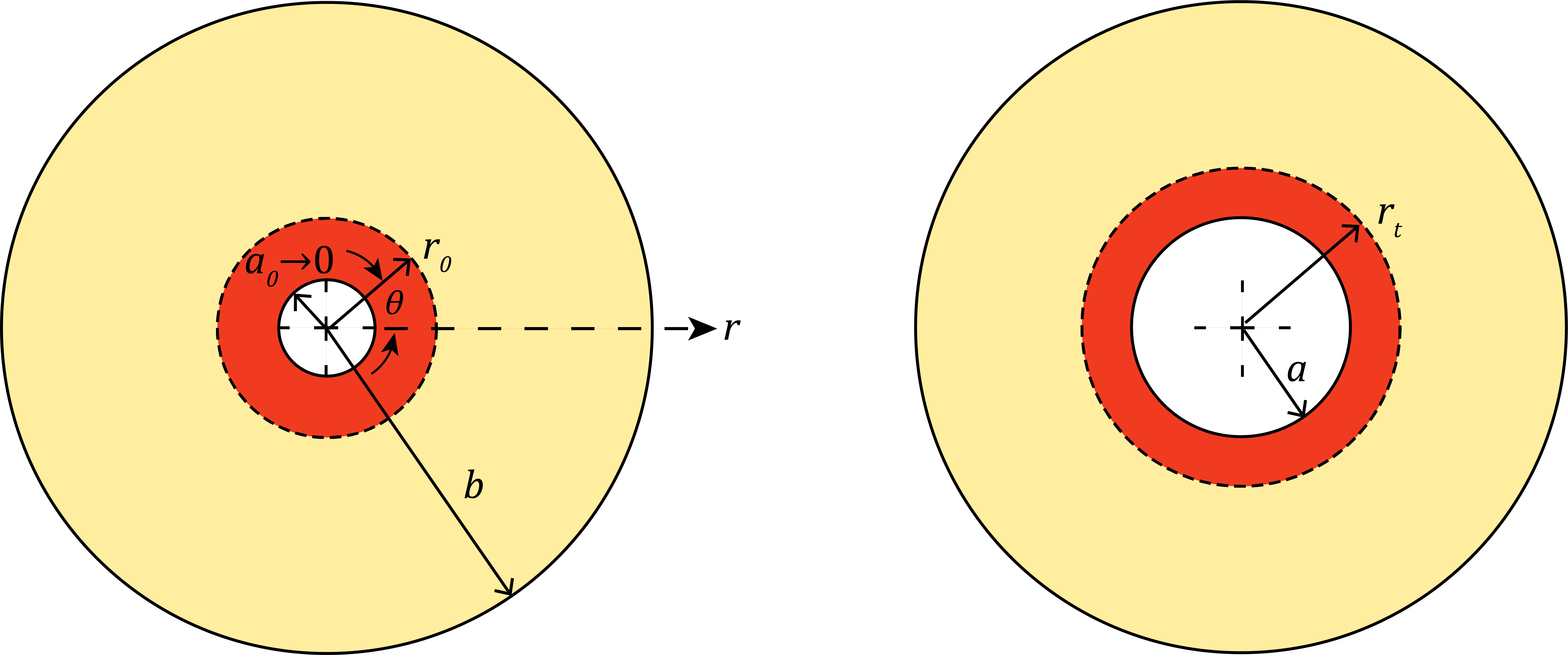}
    \caption{Schematic illustration of the plain-strain cavity expansion problem used to generate representative stress and strain histories for purposes of regression.}
    \label{fig:cavity}
\end{figure}

We now describe the function $F_0$ that returns the optimal value $\mathcal{Y}_0$ of the mesoscopic model parameters from given micromechanical model parameters $\mathcal{X}_0$. We understand optimality in the sense of achieving the closest agreement between the mesomechanical and the micromechanical models over selected deformation histories.

In order to generate histories for purposes of regression, we exercise the micromechanical and mesomechanical Johnson-Cook models under conditions corresponding to a plane-strain cavity expansion model, Fig.~\ref{fig:cavity}. Expanding cavity solutions have been commonly used to approximate conditions that arise in ballistic penetration \citep{bishop1945theory, Hanagud:1971, Forrestal:1988}. We specifically consider the expansion of a solid cylinder of radius $b$ at time $t = 0$ to a hollow cylinder of internal radius $a(t) = ct$ at time $t$ at constant rate $c$. Following \citet{bishop1945theory} we assume that the material is incompressible, whereupon the logarithmic strains in the cylindrical coordinate system $(r,\theta)$ follow as
\begin{equation}
   \strain_{rr}(r,t) = -\strain_{\theta\theta}(r,t) = \text{ln} \frac{r}{\sqrt{r^2+a^2(t)}}   \quad \text{and} \quad \strain_{zz}(r,t) = 0,
\label{eq:cavity_boundary}
\end{equation}
and the logarithmic strain rate as
\begin{equation}
    \dot{\strain}_{rr}(r,t) = -\frac{ca}{r^2+a^2(t)}.
\end{equation}
A straightforward calculation gives the Mises equivalent stress from the Johnson-Cook solution as
\begin{equation}
    \sigma_\text{eq}^\text{JC}(r,t)
    =
    \begin{cases}
        \sqrt{\frac{2}{3}}E \, \text{ln}(\frac{a^2(t)+r^2}{r^2}),
        & \sigma_\text{eq} \leq \sigma_0(r,t), \\
        \Big(A+B(\sqrt{\frac{2}{3}}\text{ln}(\frac{a^2(t)+r^2}{r^2})\Big)^n \,
        \Big(1+C\text{ln}(\frac{2ca}{\sqrt{3}\dot{\strain}_{p0}(r^2+a^2(t))})\Big),
        & \text{otherwise},
    \end{cases}
\end{equation}
where
\begin{equation}
    \sigma_0(r,t)
    =
    A\,\Big(1+C\text{ln}(\frac{2ca}{\sqrt{3}\dot{\strain}_{p0}(r^2+a^2(t))})\Big) ,
\end{equation}
is the Mises effective stress at first yield and $E$ is the Young's modulus. We then measure the discrepancy between the micromechanical and Johnson-Cook models in the root-mean square (RMS) sense
\begin{equation}\label{s67c9Z}
    {\rm Error}(\mathcal{Y}_0^{\rm trial} | \mathcal{X}_0)
    =
    \Big(
        \sum_c \int_0^b \int_0^T
            |\sigma_\text{eq}(r,t) - \sigma_\text{eq}^\text{JC}(r,t) |^2
        \, dt \, dr
    \Big)^{1/2},
\end{equation}
where the sum is carried out over a representative sample of expansion rates $c$, $\sigma_\text{eq}^\text{JC}(r,t)$ is the Mises equivalent stress computed using Johnson-Cook with parameters $\mathcal{Y}_0^{\rm trial}$ and $\sigma_\text{eq}(r,t)$ is the target Mises equivalent stress computed using the micromechanical model with parameters $\mathcal{X}_0$. The optimal Johnson-Cook parameters $\mathcal{Y}_0$ then follow by minimization of ${\rm Error}(\cdot | \mathcal{X}_0)$, i.~e.,
\begin{equation}
    \mathcal{Y}_0
    =
    {\rm argmin} \, {\rm Error}(\cdot | \mathcal{X}_0) .
\end{equation}
This step concludes the regression algorithm and the definition of the sought micro-to-mesomechanical mapping $F_0 : \mathcal{X}_0 \to \mathcal{Y}_0$.

\begin{figure}[!ht]
    \centering
    \includegraphics[width=4.0in]{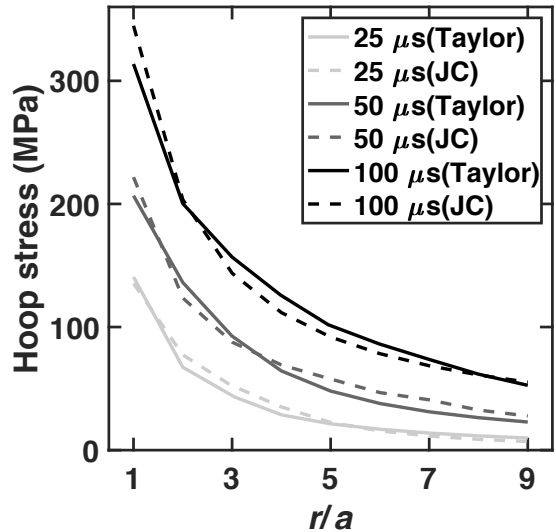}
    \caption{Comparison between the radial distributions of hoop stress $\sigma_{\theta\theta}$ computed from the micromechanical (Taylor) and the Johnson-Cook (JC) calculations.}
    \label{fig:fitting}
\end{figure}

In calculations, we employ generic algorithms (GA) \citep{mitchell1998introduction} to solve the minimization problem. The integrals in (\ref{s67c9Z}) are computed by discretizing the spatial and temporal spaces. We specifically consider a cylinder radius $b = 0.1$~mm, a single expansion rate $c=100.0$~mm/s, a maximum loading time $T=100.0~\mu$s and set the Young's modulus to $27.0$~GPa and the reference strain rate $\dot{\epsilon}_{p0} =  1.0$~s$^{-1}$. Fig.~\ref{fig:fitting} illustrates the good agreement that is achieved between the micromechanical and the Johnson-Cook model with $A=20.98$~MPa, $B=161.84$~MPa, $n=0.346$ and $C=0.430$. The maximum logarithmic strain attained in the calculations is $0.35$ and the maximum logarithmic strain rate is $5,000.0$~s$^{-1}$, which provide adequate coverage of the conditions that arise in the ballistic calculations.

\subsection{Meso-macromechanical forward solver}

\begin{figure}[!ht]
    \centering
    \includegraphics[width=6.0in]{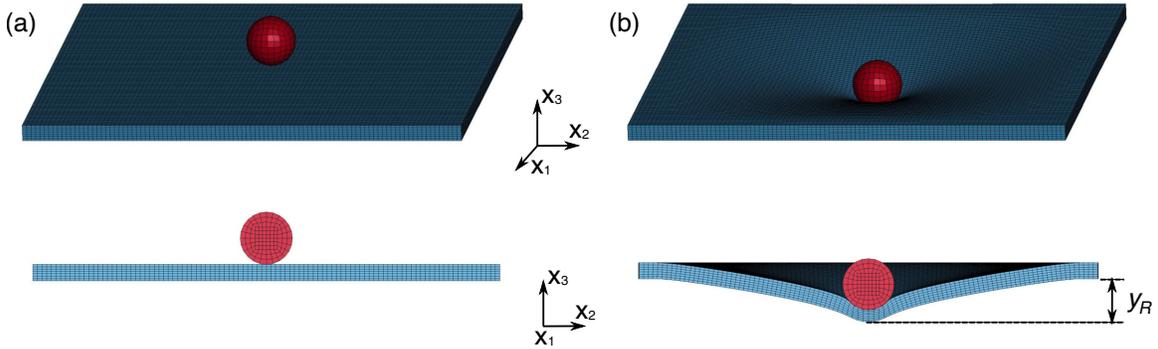}
    \caption{Schematic illustration of magnesium plate struck by a spherical steel projectile at a sub-ballistic speed. (a) Initial setup. (b) Dynamic indentation process with maximum back surface deflection labeled as $\mathcal{Y}_R$. In each subfigure, the top figure shows a perspective view of the projectile/plate system, and the bottom figure shows the cross-sectional view.}
    \label{fig:lsdyna_setup}
\end{figure}

For purposes of illustration, we consider the problem of assessing the sub-ballistic performance of a magnesium plate struck by a spherical steel projectile. The problem is solved using the explicit dynamics solver available within the commercial software LS-DYNA~\citep{hallquist2007ls} and the Johnson-Cook material model. For definiteness, we choose the maximum back surface deflection as the outcome quantity of interest $\mathcal{Y}_R$. For every sample Johson-Cook parameter set $\mathcal{X}_1$ the calculations return one value of $\mathcal{Y}_R$ and thus define a mapping
\begin{equation}
    \mathcal{Y}_R = F_1(\mathcal{X}_1) ,
\end{equation}
which completes the hierarchical model (\ref{RF3BSm}).

\begin{table}[!ht]
\centering
\caption{Fixed material parameters used in the LS-DYNA simulation. The Gruneisen parameters of the magnesium are provided by \citet{feng2017numerical}.}
\begin{tabular}{l l l l}
\hline
\hline
& Parameter & Value & Unit                  \\
\hline
\multirow{7}{*}{Target (magnesium)}
& Mass density & $1.77$ & g/$\text{cm}^3$ \\
& Young's modulus & $27.0$ & GPa \\
& Poisson's ratio & $0.35$ & - \\
& Specific heat & $1.04$ & J/(K$\cdot$g) \\
& Gruneisen intercept & $4520.0$ & m/s \\
& Gruneisen gamma & $1.54$ & -    \\
& Gruneisen slope & $1.242$ & -  \\
\hline
\multirow{3}{*}{Projectile (steel)}
& Mass density & $7.83$ & g/$\text{cm}^3$ \\
& Young's modulus & $210.0$ & GPa \\
& Poisson's ratio & $0.30$ & - \\
\hline
\end{tabular}
\label{tab:fixedparam}
\end{table}

A schematic of the finite-element model is shown in Fig.~\ref{fig:lsdyna_setup}.
The diameter of the projectile is $1.12$~cm, and the size of the plate is $10\times10\times0.35$~cm. The attack velocity is $200$~m/s with normal impact. The backface nodes of the target near the edges are fully constrained to prevent displacement in all directions. The projectile is resolved using $864$ elements, while the number of elements for the plate is $70,000$. All the elements are linear hex with single-point integration. The time-step size is adaptive and determined by the critical size of elements, with all simulations running for $500.0~\mu\text{s}$ before termination. This simulation duration is sufficiently long to allow for the rebound and separation of the projectile from the plate in all the calculations. The calculations are adiabatic with the initial temperature set at room temperature. The equation-of-state, which controls the volumetric response of the material, is assumed to be of the Gruneisen type. For simplicity, the projectile is assumed to be rigid and uncertainty-free. All other material parameters are fixed and listed in Table~\ref{tab:fixedparam}.

\subsection{Implementation of Hierarchical UQ}

\begin{table}[!ht]
\centering
\caption{Experimentally reported CRSS at room temperature for slip and twin systems in pure magnesium single crystals. The unit is MPa.}
\begin{tabular}{l l l}
\hline
\hline
& CRSS & Source \\
\hline
\multirow{6}{*}{Basal $\langle a \rangle$}
& $[0.44~0.58]$ & \cite{yoshinaga1964nonbasal} \\
& $0.48$ & \cite{akhtar1969solid} \\
& $0.52$ & \cite{herring1973thermally} \\
& $0.76$ & \cite{burke1952plastic} \\
& $0.81$ & \cite{schmid1931beitrage} \\
& $4.16$ & \cite{chapuis2011temperature} \\
\hline
\multirow{3}{*}{Prismatic $\langle a \rangle$}
& $17.0$ & \cite{yoshinaga1964nonbasal} \\
& $39.0$ & \cite{reed1957deformation} \\
& $50.0$ & \cite{flynn1961thermally} \\
\hline
\multirow{4}{*}{Pyramidal $\langle c+a \rangle$}
& $[27.76~49.22]$ & \cite{obara19731122} \\
& $44.0$ & \cite{byer2010microcompression} \\
& $[52.87~58.52]$ & \cite{ando2007temperature} \\
& $[55.48~86.0]$ & \cite{kitahara2007deformation} \\
\hline
\multirow{2}{*}{Tensile twin}
& $1.86$ & \cite{roberts1960magnesium} \\
& $[3.0,11.7]$ &  \cite{chapuis2011temperature} \\
\hline
\end{tabular}
\label{tab:CRSSexp}
\end{table}

\begin{table}[!ht]
\centering
\caption{Lower and upper bounds for the CRSS in different slip and twin systems. The unit is MPa.}
\begin{tabular}{l l l l}
\hline
\hline
Basal $\langle a \rangle$ & Prismatic $\langle a \rangle$ & Pyramidal $\langle c+a \rangle$ & Tensile twin \\
\hline
$[0.44~4.16]$ & $[17.0~50.0]$ & $[27.76~86.0]$ & $[1.86~11.7]$ \\
\hline
\end{tabular}
\label{tab:CRSSbound}
\end{table}

As already mentioned, for purposes of illustration we assume that all uncertainty in the single-crystal parameters arises from imperfect knowledge of the critical resolved shear stresses (CRSSs) in the slip and twin systems, i.~e., basal $\langle a \rangle$, prismatic $\langle a \rangle$, pyramidal $\langle c+a \rangle$ and tensile twin cf.~Fig.~\ref{fig:sliptwin}. These CRSSs are allowed to vary over a certain range in order to cover the experimental data. Table~\ref{tab:CRSSexp} lists a compilation of CRSS values reported in the literature for the slip and twin systems in pure magnesium single crystals at room temperature. Based on these data, the ranges of CRSSs used as the input in the multiscale UQ analysis are listed in Table~\ref{tab:CRSSbound}. All other material parameters in the crystal-plasticity model are fixed and listed in Table~\ref{tab:param_crystal}.

The calculation of the diameters $D_{ij}^{(k)}$ requires the solution of a global, constrained optimization problem, while the calculation of hyper-rectangles $B_k$ entails the solutions of two global optimization problems with regard to each output. In order to solve these optimization problems efficiently, we have developed a non-intrusive, high-performance computational framework based on DAKOTA Version $6.12$ software package~\citep{adams2020dakota} of the Sandia National Laboratories.
We employ genetic algorithms (GA) to solve all the optimization problems involved in the UQ analysis \citep{mitchell1998introduction, sun2020rigorous}. GA, as a global and derivative-free optimization method, offers great flexibility in applications, such as considered here, to highly non-linear non-convex problems without the availability of gradients. Another advantage of GA is its high degree of concurrency, since individuals in each iteration can be evaluated independently across multiple processors. In all the GA calculations, we choose throughout a fixed population size of $64$. The crossover rate and mutation rate are fixed at $0.8$ and $0.1$.

\subsection{Results and discussion}

We recall that the modular UQ analysis is a divide-and-conquer approach that enables each subsystem to be assessed independently. In keeping with this paradigm, we analyze the micro-meso and meso-macro maps independently and combine the results to determine integral uncertainty bounds for the entire system.

\begin{figure}[!ht]
\centering
\subfloat[ ]{\includegraphics[width = 3.0in]{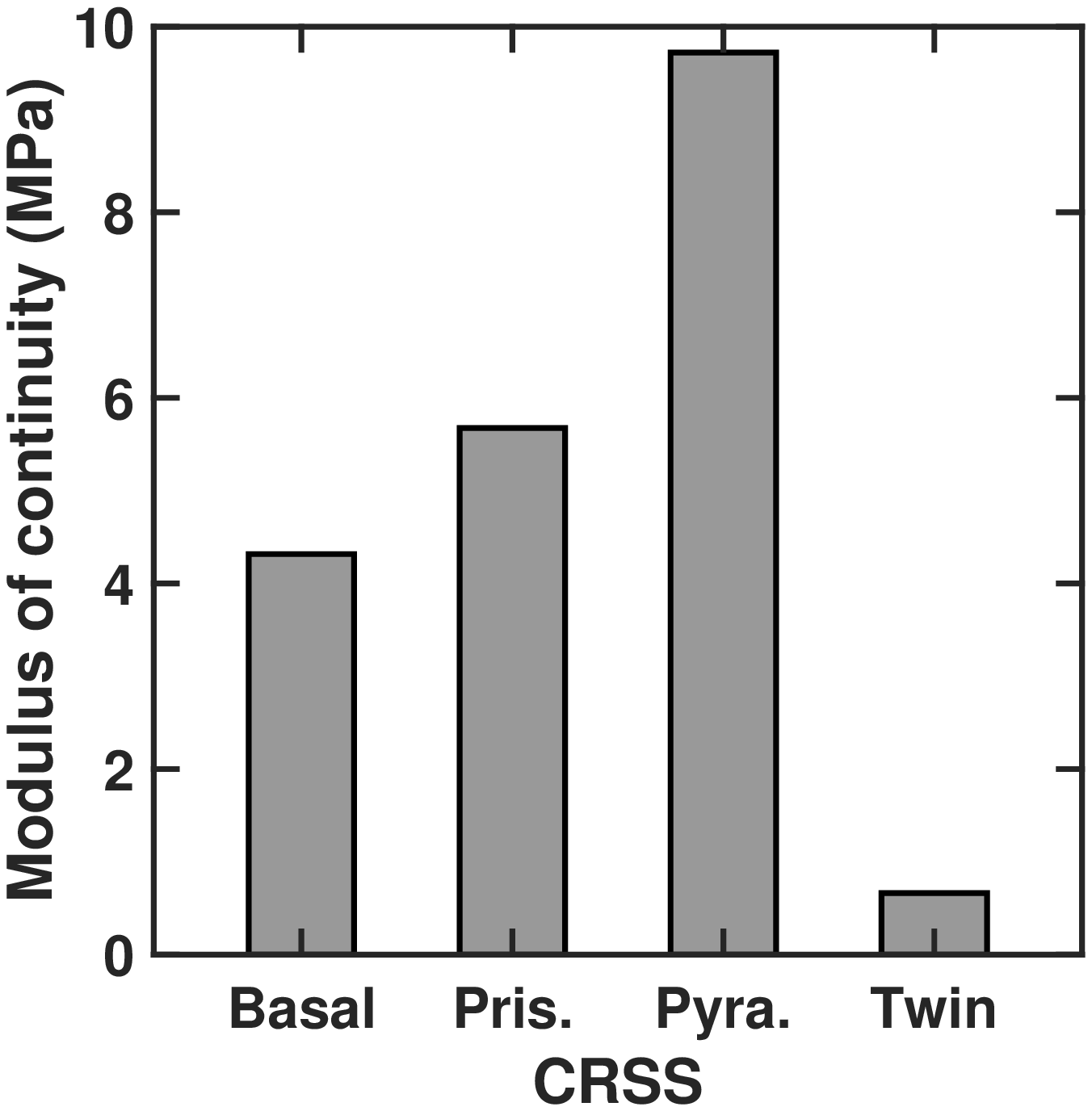}}
\subfloat[ ]{\includegraphics[width = 3.0in]{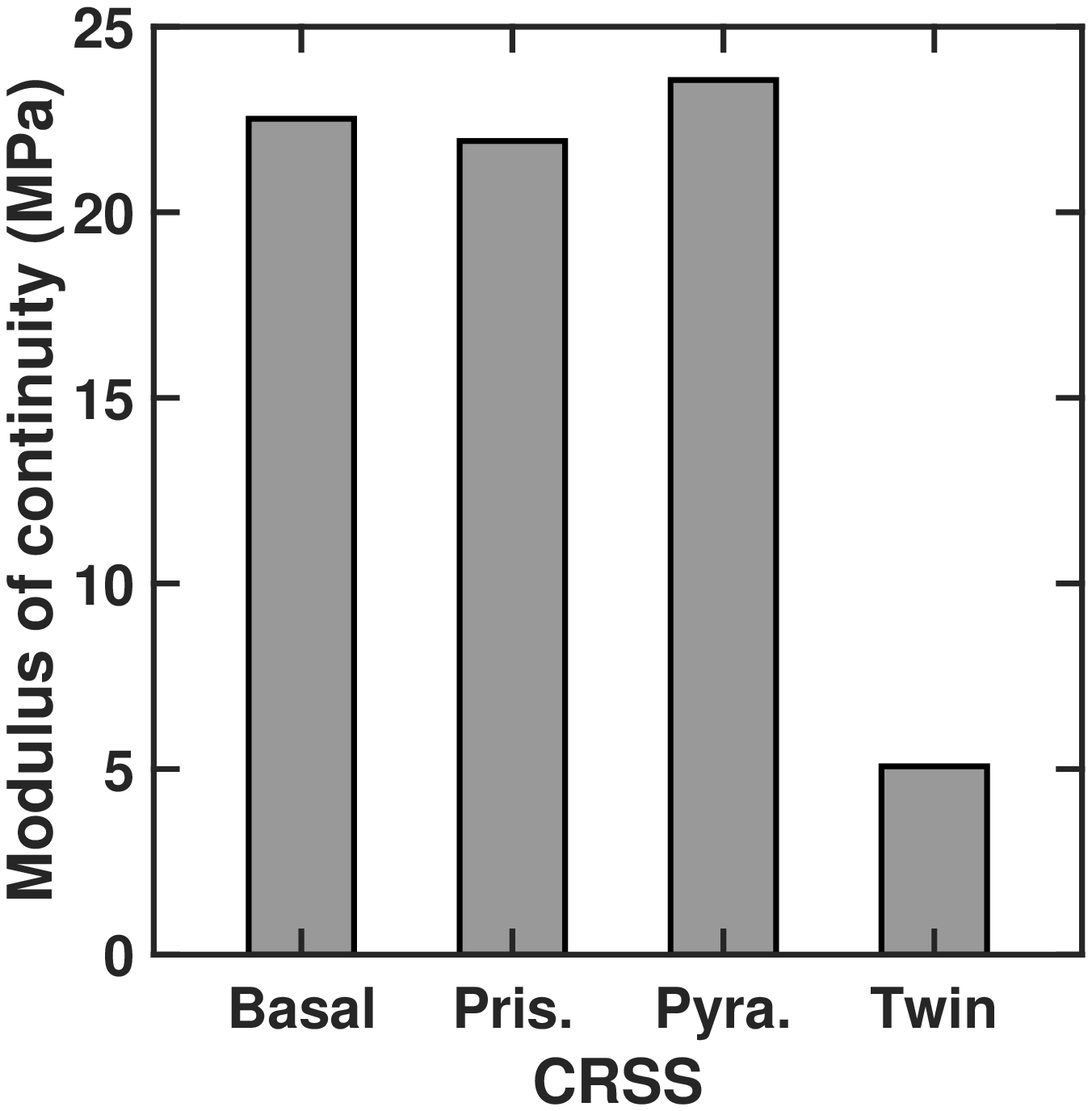}} \\
\subfloat[ ]{\includegraphics[width = 3.0in]{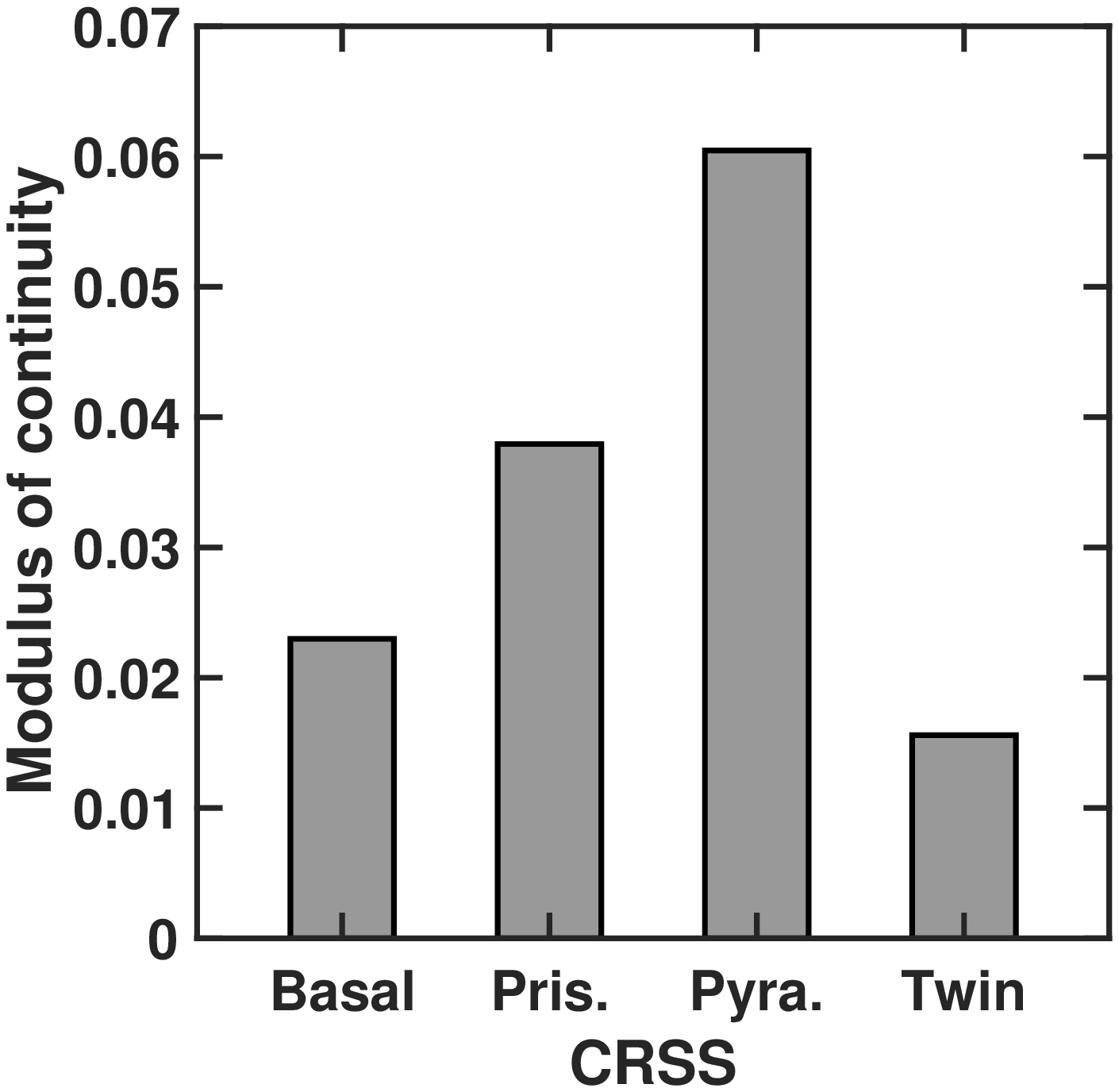}}
\subfloat[ ]{\includegraphics[width = 3.0in]{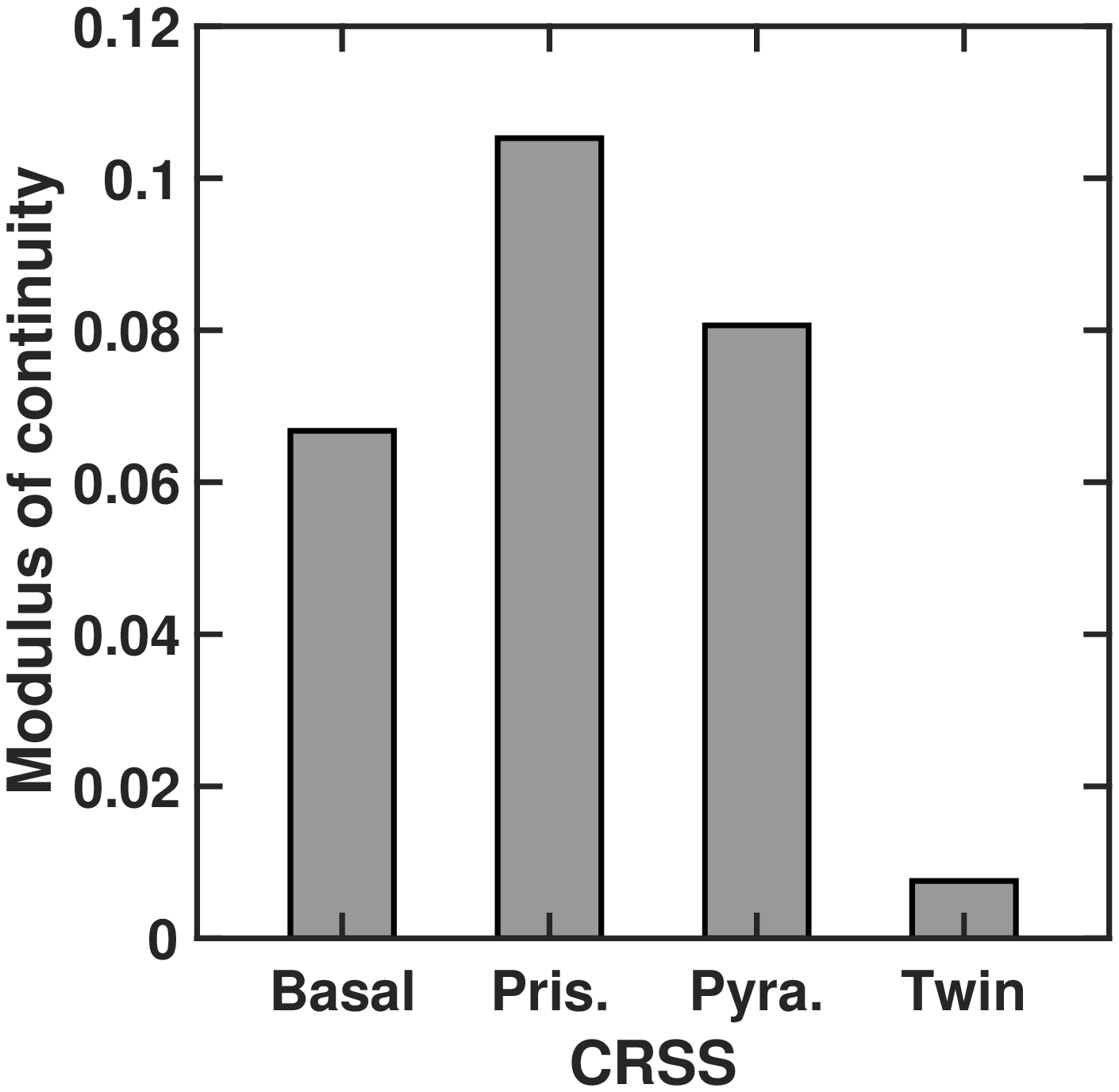}}
\caption{Moduli of continuity from single-crystal properties to Johnson-Cook (JC) parameters. (a) JC parameter $A$. (b) JC parameter $B$. (c) JC parameter $n$. (d) JC parameter $C$.}
\label{fig:crystal_JC}
\end{figure}

The moduli of continuity for the micro-meso mapping, relating uncertainties in the basal, prismatic, pyramidal and twin CRSS to uncertainties in the Johnson-Cook parameters are shown in Fig.~\ref{fig:crystal_JC}. Clearly the micro-mechanical parameters contribute to different degrees to the uncertainties in the Johnson-Cook parameters. Remarkably, twinning uncertainty contributes the least. The parameters $A$ and $n$ are most sensitive to the pyramidal CRSS, while $C$ is most sensitive to the prismatic CRSS. By contrast, the parameter $B$ is equally sensitive to all three slip mechanisms.

\begin{figure}[!ht]
    \centering
    \includegraphics[width=4.0in]{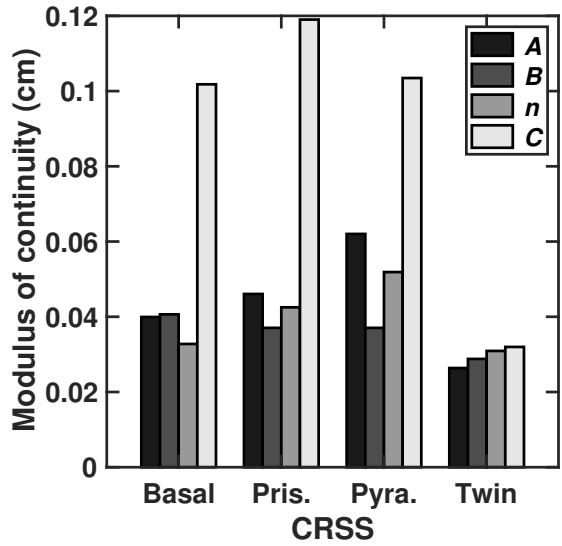}
    \caption{Moduli of continuity from single crystal properties to ballistic performance through different Johnson-Cook parameters.}
    \label{fig:cystal_ballistic}
\end{figure}

The integral moduli of continuity relating uncertainties in the CRSSs to the maximum back surface deflection of the plate are shown in Fig.~\ref{fig:cystal_ballistic}. An important property of the moduli of continuity is that, since they are dimensionally homogeneous, they can be compared and rank-ordered, which in turn provides a quantitative metric of the relative contributions of the input parameters to the overall uncertainty. In the present case, the ranking is $C>B>A>n$ for basal CRSS, $C>A>n>B$ for both prismatic and pyramidal CRSSs, and $C>n>B>A$ for twin CRSS. Remarkably, the slip uncertainty flow through the Johnson-Cook parameter $C$ is significantly larger than other paths, whereas the twin uncertainty is transmitted nearly uniformly by all Johnson-Cook parameters.

\begin{figure}[!ht]
    \centering
    \includegraphics[width=4.0in]{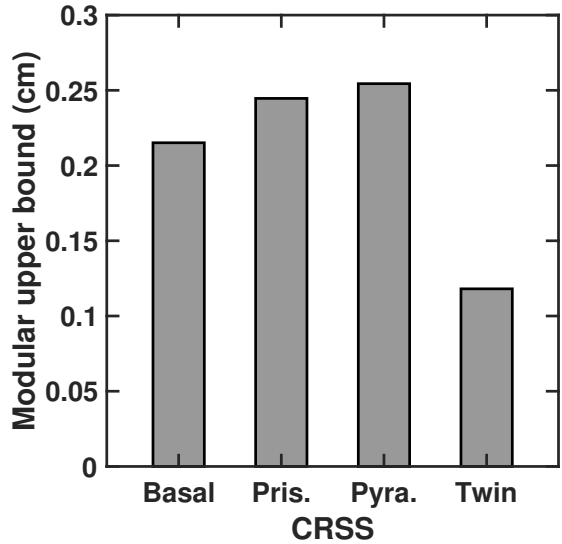}
    \caption{Sub-diameter upper bounds computed using modular approach.}
    \label{fig:cystal_ballistic_sum}
\end{figure}

Numerical results for the modular upper bounds are collected in Fig.~\ref{fig:cystal_ballistic_sum}. The rank-ordering of the CRSSs to the overall uncertainty in ballistic performance is found to be $\text{pyramidal}>\text{prismatic}>\text{basal}>\text{twin}$, with the pyramidal and prismatic CRSSs contributing the most, the twin CRSS the least and the basal CRSS in between. We also note that the modular upper bounds lie above the system diameters, cf.~Eq.~(\ref{eq:DFBound}). Thus, the modular upper bounds provide a conservative estimate of the probability of departure from the mean when inserted into McDiarmid's inequality Eq.~(\ref{eq:MicDiarmid}) in place of the system diameter.

\section{Concluding remarks}
\label{sec:concl}

We have presented a framework to assess the uncertainties of a hierarchical multi-scale material system.  The hierarchical structure of the multi-scale systems can be viewed as a directed graph, and we exploit this structure by bounding the uncertainty at each scale and then combining the partial uncertainties in a way that provides a bound on the overall or integral uncertainty.  The bound provides a conservative estimate on the integral uncertainty.  Importantly, this approach does not require integral calculations, which often are prohibitively expensive.

We have demonstrated the framework on the problem of ballistic impact of a polycrystalline magnesium plate.  Magnesium and its alloys are of current interest as promising light-weight structural and protective materials.  We start at the microscopic sub-grain scale where the behavior of various slip and twinning systems is described using extended crystal plasticity, pass to the mesoscopic scale of a representative volume involving multiple grains where the behavior is described using a Johnson-Cook constitutive model and study a macroscopic problem of ballistic impact of a plate.  We study the uncertainty of the ballistic response due to uncertainties in the strength of individual slip and twin system.

We close with a discussion of how our approach also informs the `materials-by-design' approach.  In this approach, we seek to `design' a material with desired properties by affecting some underlying mechanism at a fine scale.  As a concrete example, consider the improvement of ballistic performance (measured by maximum deflection) of a magnesium plate.  A potential way to doing so is to change the CRSS of a slip or twin system by adding solutes or precipitates.  We can estimate the change in CRSS of a particular system to a solute by conducting molecular dynamics simulations.  The question then, is how does this change in CRSS manifest itself in a change of ballistic performance.  In the notation of (\ref{RF3BSm}), we seek to understand how the change of $\mathcal{X}_0$ affects $\mathcal{Y}_R$, or the modulus of continuity of this composite map.  The method we have presented provides an outer bound (optimistic in the context of design) on this modulus.  Importantly we can compute this bound by studying individual scales ($F_0$ and $F_1$) without the need for prohibitively expensive integral calculations.

\paragraph{Acknowledgement} We are grateful to Dennis Kochmann for making available to us the single-crystal and 
Taylor averaging codes used in this work to generate data. This research was sponsored by the Army Research Laboratory and was accomplished under Cooperative Agreement Number W911NF-12-2-0022. The views and conclusions contained in this document are those of the authors and should not be interpreted as representing the official policies, either expressed or implied, of the Army Research Laboratory or the U.S. Government. The U.S. Government is authorized to reproduce and distribute reprints for Government purposes notwithstanding any copyright notation herein.

\bibliographystyle{abbrvnat}
\bibliography{references}

\begin{thebibliography}{56}
\providecommand{\natexlab}[1]{#1}
\providecommand{\url}[1]{\texttt{#1}}
\expandafter\ifx\csname urlstyle\endcsname\relax
  \providecommand{\doi}[1]{doi: #1}\else
  \providecommand{\doi}{doi: \begingroup \urlstyle{rm}\Url}\fi

\bibitem[Adams et~al.(2020)Adams, Bohnhoff, Dalbey, Ebeida, Eddy, Eldred,
  Hooper, Hough, Hu, Jakeman, Khalil, Maupin, Monschke, Ridgway, Rushdi, Seidl,
  Stephens, Swiler, and Winokur]{adams2020dakota}
B.~M. Adams, W.~J. Bohnhoff, K.~R. Dalbey, M.~S. Ebeida, J.~P. Eddy, M.~S.
  Eldred, R.~W. Hooper, P.~D. Hough, K.~T. Hu, J.~D. Jakeman, M.~Khalil, K.~A.
  Maupin, J.~A. Monschke, E.~M. Ridgway, A.~A. Rushdi, D.~T. Seidl, J.~A.
  Stephens, L.~P. Swiler, and J.~G. Winokur.
\newblock Dakota, a multilevel parallel object-oriented framework for design
  optimization, parameter estimation, uncertainty quantification, and
  sensitivity analysis: version 6.12 user's manual.
\newblock \emph{Sandia National Laboratories, Tech. Rep. SAND2020-5001}, 2020.

\bibitem[Akhtar and Teghtsoonian(1969)]{akhtar1969solid}
A.~Akhtar and E.~Teghtsoonian.
\newblock Solid solution strengthening of magnesium single crystals—i
  alloying behaviour in basal slip.
\newblock \emph{Acta Metallurgica}, 17\penalty0 (11):\penalty0 1339--1349,
  1969.

\bibitem[Ando et~al.(2007)Ando, Harada, Tsushida, Kitahara, and
  Tonda]{ando2007temperature}
S.~Ando, N.~Harada, M.~Tsushida, H.~Kitahara, and H.~Tonda.
\newblock Temperature dependence of deformation behavior in magnesium and
  magnesium alloy single crystals.
\newblock In \emph{Key Engineering Materials}, volume 345, pages 101--104.
  Trans Tech Publications Ltd, 2007.

\bibitem[Bhattacharya(2003)]{bhattacharya2003microstructure}
K.~Bhattacharya.
\newblock \emph{Microstructure of martensite: why it forms and how it gives
  rise to the shape-memory effect}, volume~2.
\newblock Oxford University Press, 2003.

\bibitem[Bishop et~al.(1945)Bishop, Hill, and Mott]{bishop1945theory}
R.~Bishop, R.~Hill, and N.~Mott.
\newblock The theory of indentation and hardness tests.
\newblock \emph{Proceedings of the Physical Society}, 57\penalty0 (3):\penalty0
  147, 1945.

\bibitem[Burke and Hibbard(1952)]{burke1952plastic}
E.~Burke and W.~Hibbard.
\newblock Plastic deformation of magnesium single crystals.
\newblock \emph{Jom}, 4\penalty0 (3):\penalty0 295--303, 1952.

\bibitem[Byer et~al.(2010)Byer, Li, Cao, and Ramesh]{byer2010microcompression}
C.~M. Byer, B.~Li, B.~Cao, and K.~Ramesh.
\newblock Microcompression of single-crystal magnesium.
\newblock \emph{Scripta Materialia}, 62\penalty0 (8):\penalty0 536--539, 2010.

\bibitem[Chang and Kochmann(2015)]{chang2015variational}
Y.~Chang and D.~M. Kochmann.
\newblock A variational constitutive model for slip-twinning interactions in
  hcp metals: application to single-and polycrystalline magnesium.
\newblock \emph{International Journal of Plasticity}, 73:\penalty0 39--61,
  2015.

\bibitem[Chapuis and Driver(2011)]{chapuis2011temperature}
A.~Chapuis and J.~H. Driver.
\newblock Temperature dependency of slip and twinning in plane strain
  compressed magnesium single crystals.
\newblock \emph{Acta Materialia}, 59\penalty0 (5):\penalty0 1986--1994, 2011.

\bibitem[Dashti and Stuart(2011)]{dashti2011uncertainty}
M.~Dashti and A.~M. Stuart.
\newblock Uncertainty quantification and weak approximation of an elliptic
  inverse problem.
\newblock \emph{SIAM Journal on Numerical Analysis}, 49\penalty0 (6):\penalty0
  2524--2542, 2011.

\bibitem[Doob(1940)]{doob1940regularity}
J.~L. Doob.
\newblock Regularity properties of certain families of chance variables.
\newblock \emph{Transactions of the American Mathematical Society}, 47\penalty0
  (3):\penalty0 455--486, 1940.

\bibitem[Efimov(2001)]{efimov2001modulus}
A.~Efimov.
\newblock Modulus of continuity, encyclopaedia of mathematics, 2001.

\bibitem[Feng et~al.(2017)Feng, Chen, Zhou, Dai, An, and
  Yuan]{feng2017numerical}
J.~Feng, P.~Chen, Q.~Zhou, K.~Dai, E.~An, and Y.~Yuan.
\newblock Numerical simulation of explosive welding using smoothed particle
  hydrodynamics method.
\newblock \emph{International Journal of Multiphysics}, 11\penalty0 (3), 2017.

\bibitem[Flynn et~al.(1961)Flynn, Mote, and Dorn]{flynn1961thermally}
P.~W. Flynn, J.~Mote, and J.~E. Dorn.
\newblock On the thermally activated mechanism of prismatic slip in magnesium
  single crystals.
\newblock \emph{Transactions of the Metallurgical Society of AIME},
  221\penalty0 (6):\penalty0 1148--1154, 1961.

\bibitem[Forrestal et~al.(1988)Forrestal, Okajima, and Luk]{Forrestal:1988}
M.~J. Forrestal, K.~Okajima, and V.~K. Luk.
\newblock Penetration of 6061--t651 aluminum targets with rigid long rods.
\newblock \emph{Journal of Applied Mechanics}, 55\penalty0 (4):\penalty0
  755--760, 1988.

\bibitem[Geers et~al.(2010)Geers, Kouznetsova, and Brekelmans]{geers2010multi}
M.~G. Geers, V.~G. Kouznetsova, and W.~Brekelmans.
\newblock Multi-scale computational homogenization: Trends and challenges.
\newblock \emph{Journal of computational and applied mathematics}, 234\penalty0
  (7):\penalty0 2175--2182, 2010.

\bibitem[Green(1954)]{green1954markoff}
M.~S. Green.
\newblock Markoff random processes and the statistical mechanics of
  time-dependent phenomena. ii. irreversible processes in fluids.
\newblock \emph{The Journal of Chemical Physics}, 22\penalty0 (3):\penalty0
  398--413, 1954.

\bibitem[Haar(1933)]{haar1933massbegriff}
A.~Haar.
\newblock Der massbegriff in der theorie der kontinuierlichen gruppen.
\newblock \emph{Annals of mathematics}, pages 147--169, 1933.

\bibitem[Hall(1951)]{hall1951deformation}
E.~Hall.
\newblock The deformation and ageing of mild steel: Iii discussion of results.
\newblock \emph{Proceedings of the Physical Society. Section B}, 64\penalty0
  (9):\penalty0 747, 1951.

\bibitem[Hallquist et~al.(2007)]{hallquist2007ls}
J.~O. Hallquist et~al.
\newblock Ls-dyna keyword users manual.
\newblock \emph{Livermore Software Technology Corporation}, 970:\penalty0
  299--800, 2007.

\bibitem[Hanagud and Ross(1971)]{Hanagud:1971}
S.~Hanagud and B.~Ross.
\newblock Large deformation, deep penetration theory for a compressible
  strain-hardening target material.
\newblock \emph{AIAA Journal}, 9:\penalty0 905--911, 1971.

\bibitem[Herring and Meshii(1973)]{herring1973thermally}
R.~Herring and M.~Meshii.
\newblock Thermally activated deformation of gold single crystals.
\newblock \emph{Metallurgical Transactions}, 4\penalty0 (9):\penalty0
  2109--2114, 1973.

\bibitem[Hutchinson(1970)]{hutchinson1970elastic}
J.~Hutchinson.
\newblock Elastic-plastic behaviour of polycrystalline metals and composites.
\newblock \emph{Proceedings of the Royal Society of London. A. Mathematical and
  Physical Sciences}, 319\penalty0 (1537):\penalty0 247--272, 1970.

\bibitem[Johnson(1983)]{johnson1983constitutive}
G.~R. Johnson.
\newblock A constitutive model and data for materials subjected to large
  strains, high strain rates, and high temperatures.
\newblock \emph{Proc. 7th Inf. Sympo. Ballistics}, pages 541--547, 1983.

\bibitem[Kang et~al.(2012)Kang, Bulatov, and Cai]{kang2012singular}
K.~Kang, V.~V. Bulatov, and W.~Cai.
\newblock Singular orientations and faceted motion of dislocations in
  body-centered cubic crystals.
\newblock \emph{Proceedings of the National Academy of Sciences}, 109\penalty0
  (38):\penalty0 15174--15178, 2012.

\bibitem[Kitahara et~al.(2007)Kitahara, Ando, Tsushida, Kitahara, and
  Tonda]{kitahara2007deformation}
T.~Kitahara, S.~Ando, M.~Tsushida, H.~Kitahara, and H.~Tonda.
\newblock Deformation behavior of magnesium single crystals in c-axis
  compression.
\newblock In \emph{Key Engineering Materials}, volume 345, pages 129--132.
  Trans Tech Publ, 2007.

\bibitem[Kohler et~al.(2004)Kohler, Kizler, and Schmauder]{kohler2004atomistic}
C.~Kohler, P.~Kizler, and S.~Schmauder.
\newblock Atomistic simulation of precipitation hardening in $\alpha$-iron:
  influence of precipitate shape and chemical composition.
\newblock \emph{Modelling and Simulation in Materials Science and Engineering},
  13\penalty0 (1):\penalty0 35, 2004.

\bibitem[Koslowski et~al.(2002)Koslowski, Cuitino, and Ortiz]{koslowski2002a}
M.~Koslowski, A.~Cuitino, and M.~Ortiz.
\newblock A phase-field theory of dislocation dynamics, strain hardening and
  hysteresis in ductile single crystals.
\newblock \emph{Journal of the Mechanics and Physics of Solids}, 50:\penalty0
  2597–2635, 2002.

\bibitem[Kubo(1957)]{kubo1957statistical}
R.~Kubo.
\newblock Statistical-mechanical theory of irreversible processes. i. general
  theory and simple applications to magnetic and conduction problems.
\newblock \emph{Journal of the Physical Society of Japan}, 12\penalty0
  (6):\penalty0 570--586, 1957.

\bibitem[Leibfried and Breuer(2006)]{leibfried2006point}
G.~Leibfried and N.~Breuer.
\newblock \emph{Point defects in metals I: introduction to the theory},
  volume~81.
\newblock Springer, 2006.

\bibitem[Li et~al.(2010)Li, Habbal, and Ortiz]{li2010a}
B.~Li, F.~Habbal, and M.~Ortiz.
\newblock Optimal transportation meshfree approximation schemes for fluid and
  plastic flows.
\newblock \emph{International Journal for Numerical Methods in Engineering},
  83\penalty0 (12):\penalty0 1541--1579, 2010.

\bibitem[Lucas et~al.(2008)Lucas, Owhadi, and Ortiz]{lucas2008rigorous}
L.~J. Lucas, H.~Owhadi, and M.~Ortiz.
\newblock Rigorous verification, validation, uncertainty quantification and
  certification through concentration-of-measure inequalities.
\newblock \emph{Computer Methods in Applied Mechanics and Engineering},
  197\penalty0 (51-52):\penalty0 4591--4609, 2008.

\bibitem[Messerschmidt(2010)]{messerschmidt2010dislocation}
U.~Messerschmidt.
\newblock \emph{Dislocation dynamics during plastic deformation}, volume 129.
\newblock Springer Science \& Business Media, 2010.

\bibitem[Miljacic et~al.(2011)Miljacic, Demers, and van~de
  Walle]{Miljacic:2011}
L.~Miljacic, S.~Demers, and A.~van~de Walle.
\newblock Equation of state and viscosity of tantalum and iron from first
  principles.
\newblock \emph{American Physical Society}, APS March Meeting 2011:\penalty0
  21--25, March 2011.

\bibitem[Mitchell(1998)]{mitchell1998introduction}
M.~Mitchell.
\newblock \emph{An introduction to genetic algorithms}.
\newblock MIT press, 1998.

\bibitem[Mukasey et~al.(2008)Mukasey, Sedgwick, and Hagy]{standard20080101}
M.~Mukasey, J.~L. Sedgwick, and D.~Hagy.
\newblock Ballistic resistance of body armor, nij standard-0101.06.
\newblock \emph{US Department of Justice (www. ojp. usdoj. gov/nij)}, 2008.

\bibitem[Obara et~al.(1973)Obara, Yoshinga, and Morozumi]{obara19731122}
T.~Obara, H.~Yoshinga, and S.~Morozumi.
\newblock $\{11\bar{2}2\}\langle 1123\rangle$ slip system in magnesium.
\newblock \emph{Acta Metallurgica}, 21\penalty0 (7):\penalty0 845--853, 1973.

\bibitem[Olson(2000)]{olson2000designing}
G.~B. Olson.
\newblock Designing a new material world.
\newblock \emph{Science}, 288\penalty0 (5468):\penalty0 993--998, 2000.

\bibitem[Ortiz and Repetto(1999)]{ortiz1999a}
M.~Ortiz and E.~Repetto.
\newblock Non-convex energy minimization and dislocation structures in ductile
  single crystals.
\newblock \emph{Journal of the Mechanics and Physics of Solids}, 47:\penalty0
  397–462, 1999.

\bibitem[Ortiz et~al.(2001)Ortiz, Cuiti{\~n}o, Knap, and Koslowski]{Ortiz:2001}
M.~Ortiz, A.~M. Cuiti{\~n}o, J.~Knap, and M.~Koslowski.
\newblock Mixed atomistic-continuum models of material behavior: The art of
  transcending atomistics and informing continua.
\newblock \emph{{MRS} Bulletin}, 26\penalty0 (3):\penalty0 216--221, 2001.

\bibitem[Petch(1953)]{petch1953cleavage}
N.~Petch.
\newblock The cleavage strength of polycrystals.
\newblock \emph{Journal of the Iron and Steel Institute}, 174:\penalty0 25--28,
  1953.

\bibitem[Phillips(2001)]{p_book_01}
R.~Phillips.
\newblock \emph{Crystals, defects and microstructures: Modeling across scales}.
\newblock Cambridge University Press, 2001.

\bibitem[Reed-Hill and Robertson(1957)]{reed1957deformation}
R.~E. Reed-Hill and W.~D. Robertson.
\newblock Deformation of magnesium single crystals by nonbasal slip.
\newblock \emph{Jom}, 9\penalty0 (4):\penalty0 496--502, 1957.

\bibitem[Roberts(1960)]{roberts1960magnesium}
C.~S. Roberts.
\newblock \emph{Magnesium and its Alloys}.
\newblock Wiley, 1960.

\bibitem[Schmid(1931)]{schmid1931beitrage}
E.~Schmid.
\newblock Beitr{\"a}ge zur physik und metallographie des magnesiums.
\newblock \emph{Zeitschrift f{\"u}r Elektrochemie und angewandte physikalische
  Chemie}, 37\penalty0 (8-9):\penalty0 447--459, 1931.

\bibitem[Segall et~al.(2001)Segall, Arias, Strachan, and Goddard]{Segall:2001}
D.~E. Segall, T.~A. Arias, A.~Strachan, and W.~A. Goddard.
\newblock Accurate calculations of the peierls stress in small periodic cells.
\newblock \emph{Journal of Computer-Aided Materials Design}, 8\penalty0
  (2):\penalty0 161--172, 2001.

\bibitem[Sharp and Wood-Schultz(2003)]{sharp2003qmu}
D.~H. Sharp and M.~M. Wood-Schultz.
\newblock Qmu and nuclear weapons certification-what's under the hood?
\newblock \emph{Los Alamos Science}, 28:\penalty0 47--53, 2003.

\bibitem[Steffens(2006)]{steffens2006constructive}
K.-G. Steffens.
\newblock Constructive function theory: Kharkiv.
\newblock \emph{The History of Approximation Theory: From Euler to Bernstein},
  pages 167--190, 2006.

\bibitem[Sun et~al.(2020)Sun, Kirchdoerfer, and Ortiz]{sun2020rigorous}
X.~Sun, T.~Kirchdoerfer, and M.~Ortiz.
\newblock Rigorous uncertainty quantification and design with uncertain
  material models.
\newblock \emph{International Journal of Impact Engineering}, 136:\penalty0
  103418, 2020.

\bibitem[Talagrand(1996)]{talagrand1996new}
M.~Talagrand.
\newblock A new look at independence.
\newblock \emph{The Annals of probability}, pages 1--34, 1996.

\bibitem[Taylor(1938)]{t_jim_38}
G.~Taylor.
\newblock Plastic strain in metals.
\newblock \emph{Journal of the Institute of Metals}, 62:\penalty0 307--324,
  1938.

\bibitem[Topcu et~al.(2011)Topcu, Lucas, Owhadi, and Ortiz]{topcu2011rigorous}
U.~Topcu, L.~J. Lucas, H.~Owhadi, and M.~Ortiz.
\newblock Rigorous uncertainty quantification without integral testing.
\newblock \emph{Reliability Engineering \& System Safety}, 96\penalty0
  (9):\penalty0 1085--1091, 2011.

\bibitem[Wang et~al.(2001)Wang, Strachan, Cagin, and Goddard]{Wang:2001}
G.~Wang, A.~Strachan, T.~Cagin, and W.~A. Goddard.
\newblock Molecular dynamics simulations of 1/2 a<1 1 1> screw dislocation in
  ta.
\newblock \emph{Materials Science and Engineering: A}, 309-310:\penalty0 133 --
  137, 2001.
\newblock Dislocations 2000: An International Conference on the Fundamentals of
  Plastic Deformation.

\bibitem[Wei and Anand(2004)]{wei2004grain}
Y.~Wei and L.~Anand.
\newblock Grain-boundary sliding and separation in polycrystalline metals:
  application to nanocrystalline fcc metals.
\newblock \emph{Journal of the Mechanics and Physics of Solids}, 52\penalty0
  (11):\penalty0 2587--2616, 2004.

\bibitem[Yoshinaga and Horiuchi(1964)]{yoshinaga1964nonbasal}
H.~Yoshinaga and R.~Horiuchi.
\newblock On the nonbasal slip in magnesium crystals.
\newblock \emph{Transactions of the Japan Institute of Metals}, 5\penalty0
  (1):\penalty0 14--21, 1964.

\bibitem[Zhao et~al.(2004)Zhao, Radovitzky, and Cuitino]{zhao2004a}
Z.~Zhao, R.~Radovitzky, and A.~Cuitino.
\newblock A study of surface roughening in fcc metals using direct numerical
  simulation.
\newblock \emph{Acta Materialia}, 52:\penalty0 5791–5804, 2004.

\end{thebibliography}
\end{document}